\documentclass{article}

\PassOptionsToPackage{numbers, compress}{natbib}

\usepackage[preprint]{neurips_2024}
\usepackage[utf8]{inputenc} 
\usepackage[T1]{fontenc}    
\usepackage{hyperref}       
\usepackage{url}            
\usepackage{booktabs}       
\usepackage{amsfonts}       
\usepackage{nicefrac}       
\usepackage{microtype}      
\usepackage{xcolor}         
\usepackage{cleveref}
\usepackage{xspace}
\newcommand{\sys}{\textsc{PostconEval}\xspace}
\usepackage{tikz}
\newcommand*\circled[1]{\tikz[baseline=(char.base)]{
            \node[shape=circle,draw,inner sep=.2pt] (char) {#1};}}
\usepackage{graphicx} 
\usepackage{multirow}
\usepackage{makecell}

\title{Beyond Code Generation: Assessing Code LLM Maturity with Postconditions}
\author{
  Fusen He \\
  Nanjing University
  \And
  Juan Zhai \\
  University of Massachusetts, Amherst (UMass)
  \And
  Minxue Pan \\
  Nanjing University
}
\begin{document}
\maketitle
\begin{abstract}
Most existing code Large Language Model (LLM) benchmarks, e.g., EvalPlus, focus on the code generation tasks.
Namely, they contain a natural language description of a problem and ask the LLM to write code to solve the problem.
We argue that they do not capture all capabilities needed to assess the quality of a code LLM.
In this paper, we propose a code LLM maturity model, based on the postcondition generation problem, to access a more complete set of code LLM capabilities.
We choose the postcondition generation problem as it requires the code LLM to understand the code including semantics, natural language, and also have the capability to generate unambiguous postconditions in programming languages (i.e., the generation capablity).
Moreover, postconditions have various types, requiring different levels of these capabilities, making it suitable to evaluate the maturity of the code LLM.
Based on our designed maturity model, we augment the EvalPlus dataset to a postcondition testing benchmark, and evaluated several open-sourced models.
Our results highlight the necessary improvements needed for better LLMs for code.
Code: \url{https://github.com/MatureModel/PostcondGen}
\end{abstract}

\section{Introduction}\label{sec:intro}

The advent of Large Language Models (LLMs)\cite{Radford2018ImprovingLU} has significantly transformed the landscape of natural language processing (\cite{Zhang2023PromptingLL, Yang2023ExploringTL, Hegselmann2022TabLLMFC}) and code generation. 
For example, models such as OpenAI’s GPT-3 and Codex have demonstrated remarkable proficiency in generating human-like text and writing functional code snippets from natural language prompts\cite{Chen2021EvaluatingLL}. 
The AI-powered coding assistant, GitHub Copilot has 1.3 million paid subscribers, a 30\% quarter-over-quarter increase\cite{CopilotNews}.

To understand the usefulness and limitations of code LLMs, there have been various benchmarks.
Existing benchmarks, such as EvalPlus\cite{Liu2023IsYC}, have primarily focused on assessing the ability of LLMs to generate code from natural language descriptions of problems. 
These benchmarks typically involve providing a description of a programming task and evaluating the LLM based on its ability to generate a correct and functional code snippet that solves the problem\cite{Zhao2023ASO}. 
While effective in gauging certain aspects of code generation, these benchmarks do not fully encompass the diverse capabilities required for a comprehensive evaluation of code LLMs.

Moreover, using code generation tasks to measure the maturity of code LLM, although insightful, overlooks other critical aspects of code understanding and generation capabilities that are important for a code LLM. 
For example, the understanding of data structures is a critical aspect of code LLM, which requires the understanding of data type (primitive and combined), logical reasoning, value or property understanding (which may require domain knowledge understanding), and many more.

In this paper, we propose a more holistic approach to evaluating code LLMs through a maturity model based on the postcondition generation problem. 
This approach aims to capture a broader spectrum of LLM capabilities. 
Postcondition generation requires LLMs to not only understand the semantics of the code and the accompanying natural language description but also to generate precise and unambiguous postconditions in programming languages. 
These postconditions can take various forms, each demanding different levels of understanding and generation capabilities, making them an ideal metric for evaluating the maturity and robustness of code LLMs.

Our proposed maturity model expands upon the existing EvalPlus dataset, transforming it into a benchmark that includes postcondition testing. 
This enhanced benchmark allows for a more detailed assessment of LLM performance across multiple dimensions of code understanding and generation. 
Moreover, we have also designed few-short prompting and category-based prompts that leverages the task and problem specific information to test the capablity of code LLMs.
We evaluated several open-sourced models using this augmented dataset, and our results underscore the areas where improvements are necessary to advance the capabilities of LLMs in code-related tasks.
Results have also shown the effectiveness of our benchmark in capturing the necessary code LLM capablities.

By incorporating postcondition generation into the evaluation framework, we provide a more comprehensive measure of LLM capabilities. 
This paper presents our findings and highlights the essential enhancements required to develop more proficient and versatile code LLMs.

Our contributions are threefold: \protect\circled{1} \textbf{Proposal of a Code LLM Maturity Model}: We introduce a comprehensive maturity model based on the postcondition generation problem, which evaluates a broader set of capabilities in code LLMs beyond traditional code generation tasks;
\protect\circled{2} \textbf{Augmentation of the EvalPlus Dataset}: We expand the EvalPlus dataset to include postcondition testing, creating a more robust benchmark that assesses the LLMs' ability to generate correct and diverse postconditions; and
\protect\circled{3} \textbf{Evaluation of Open-Sourced Models}: We conduct extensive evaluations of several open-sourced LLMs using our augmented dataset, providing detailed insights into their performance and highlighting areas for improvement in their code generation and understanding capabilities.
\vspace{-12pt}
\section{Related Work}\label{sec:relwk}
\vspace{-12pt}



\textbf{Large Language Model}: Large Language Models (LLMs) have received increasing attention due to their "emergent abilities" which are not presented in small-scale models~\cite{Wei2022EmergentAO}. Their understanding and generation capabilities are commonly evaluated through various tasks spanning multiple domains including Question Answering, Reasoning, Math/Science and Coding~\cite{Touvron2023Llama2O, Touvron2023LLaMAOA}. 
Some LLMs are specifically trained for coding, such as Starcoder~\cite{Li2023StarCoderMT}, CodeGen~\cite{Nijkamp2022CodeGenAO} and CodeGen2~\cite{Nijkamp2023CodeGen2LF}.
An increasing number of studies also explore LLMs' potential in software development tasks, such as code generation~\cite{Chen2021EvaluatingLL, Liu2023IsYC, Kumar2022DeepLD, Deng2022RecentAI}, code reparation~\cite{Jiang2023ImpactOC, Wang2021CodeT5IU}, test generation~\cite{Tufano2020GeneratingAA, Schfer2023AdaptiveTG, Siddiq2023UsingLL} and mutant generation\cite{Ojdani2021SyntacticVS, Ojdani2023OnCM}.
The LLM can be evaluated by “Levels of AGI” based on depth (performance) and breadth (generality) of capabilities~\cite{Morris2023LevelsOA}.
Compared to these work, we focus on postcondition generation problem to further assess LLMs capabilities. 


\textbf{Formal Specification Generation}: There has been a long line of research on formal specification generation. 
They can be broadly categorized into two types: code-based generation and natural language-based generation. 
Code-based generation are those generating formal specification from existing code. 
Most of these work use static analysis~\cite{Chen2016SupportingOC, Shoham2007StaticSM, Flanagan2001HoudiniAA}, and dynamic analysis such as DIDUCE system~\cite{Hangal2002TrackingDS} and Daikon system~\cite{Ernst2007TheDS} which generate formal specifications by capturing a program's behaviors during execution. 
Machine learning-based approaches are increasingly explored to generate specifications from code, such as Code2Inv~\cite{Si2020Code2InvAD} and CLN2INV~\cite{Ryan2019CLN2INVLL}. 
Molina et al. also presented a learning technique to construct data structure invariants based on artificial neural network~\cite{Molina2019TrainingBC}. 
Additionally, many recent work also leverage LLMs to generate formal specifications based on the program implementation. 
Ma et al. assessed LLMs capability on generating JML style specifications for Java programs~\cite{Ma2024SpecGenAG}. 
Pei et al. finetuned language models to generate program invariants~\cite{Pei2023CanLL}. 
Janssen et al. use ChatGPT to generated loop invariants to support program verification~\cite{Janssen2023CanCS}. 
Unlike these work, our work belongs to natural language-based generation, which aims to generate formal specifications from natural languagev(NL), such as documents and code comments. 
The most common techniques are heuristics, using predefined rules to analyze the program properties from natural language text~\cite{Pandita2012InferringMS, BlasiGKGEPC2018, Tan2007icommentBO, Tan2011aCommentMA, Tan2012tCommentTJ}. 
There are also some efforts on exploring LLMs to produce specifications from NL.
Endres et al. generated postconditions from NL with LLM to formalize program intents.~\cite{Endres2023FormalizingNL}. 
Compared to these work, our approach delves deeper into the maturity level of LLMs by assessing their capabilities on generating different types of postconditions.

\section{Postcondition \& Taxonomy}\label{sec:pre}
\vspace{-4pt}

\begin{figure}
    \centering
        \footnotesize
        \includegraphics[width=\linewidth]{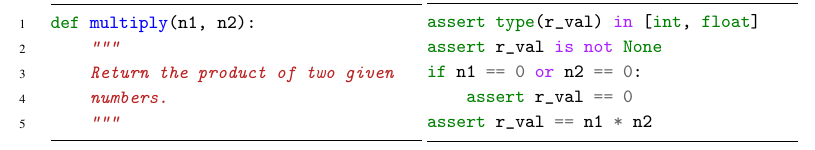}
        \caption{An Example for Postconditions Generated by LLMs}   
        \vspace{-11pt}
        \label{fig:postcondition_example}
\end{figure}

\noindent\textbf{Postcondition.}
Formal specification is critical to the correctness and safety of programs. 
It conveys programmers' intentions and defines expected behaviors by providing a precise and unambiguous description written in formal languages.
Depending on the location in the program, common specifications include pre- and post-conditions, and loop invariants\cite{Pearce2018AnIT}.
Postcondition is an essential and most popular form of specifications.
It is a condition that always holds true after the execution of a given code snippet, assuming that all preconditions are satisfied.
An example is shown in \autoref{fig:postcondition_example}. The \texttt{multiply} function has multiple postconditions, checking different aspects of the program, such as \texttt{assert assert r\_val == n1 * n2} where \texttt{r\_val} is the returned value.

Postcondition generation is an effective metric for evaluating the maturity of code LLMs because it encompasses a comprehensive range of capabilities required for robust code understanding and generation. 
Postconditions require the model to understand the semantics of the code, accurately interpret natural language descriptions, and generate precise, unambiguous conditions that must hold true after code execution. 
This task tests the model’s proficiency in logical reasoning, domain knowledge, and syntactic correctness. 
By assessing an LLM’s ability to generate diverse postconditions, including type checks, boundary conditions, and complex logical assertions, we gain a holistic view of its strengths and limitations in handling real-world coding scenarios. 
This multifaceted approach ensures that the model is not only capable of generating syntactically correct code but also understands the underlying logic and requirements, thereby providing a reliable measure of its overall maturity and effectiveness. \looseness=-1

\noindent\textbf{Postcondition Taxonomy.}
Based on verification objectives and data types, postcoditions can be categorized as ten types, as shown in \autoref{tab:taxonomy}.
Every type of postconditions verifies different aspects of the program. 
In particular, \textit{type}, \textit{NULL}, \textit{boundary}, and \textit{equality} postconditions can be applied to all variables. 
Others including \textit{arithmetic bounds}, \textit{boolean condition}, \textit{string format}, and \textit{container elements/properties} postcondtions can only be applied to specific variable types according to their definitions.
We classify all other types of postconditions (i.e., reflecting concrete logic) in the other category. \looseness=-1





\begin{table}
  \centering
  \scriptsize
  \caption{Postcondition Taxonomy Based on Check Objectives and Data Types (\texttt{assert} omitted)}\label{tab:taxonomy}
  \begin{tabular}{
  >{\raggedleft\arraybackslash}p{2.8cm}
  >{\raggedright\arraybackslash}p{6.6cm}
  >{\raggedright\arraybackslash}p{3.2cm}
  }
  \toprule
  Category                  & Explanation & Examples \\ \midrule
  Type                & Check whether the type of variable is correct  &  \begin{minipage}[t]{2.8cm}
\texttt{isinstance(var, str)}
\end{minipage} \\ 
  NULL                & Check whether variable is NULL & \begin{minipage}[t]{2.8cm}
\texttt{var is None}
\end{minipage} \\ 
  Boundary       & Check whether variable is a boundary case & \begin{minipage}[t]{2.8cm}
\texttt{age == 0}
\end{minipage} \\ 
  Equality            & Check if the value of a variable is expected &  \begin{minipage}[t]{2.8cm}
\texttt{pi == 3.14}
\end{minipage} \\ 
  Arithmetic Bounds & Check if a numerical value is in expected range  &  \begin{minipage}[t]{2.8cm}
\texttt{0 <= var <= 100}
\end{minipage} \\ 
  Boolean Condition & Check if a Boolean variable value is correct &  \begin{minipage}[t]{2.8cm}
\texttt{leap==True if y==2000}
\end{minipage} \\ 
  String Format       & Check if a string's format is expected &  \begin{minipage}[t]{2.8cm}
\texttt{s.startswith('a')}
\end{minipage} \\ 
  Container Element & Check elements of a container are correct &   \begin{minipage}[t]{2.8cm}
\texttt{x!=0 for x in [1,2]}
\end{minipage} \\ 
  Container Property  & Check if container variable's properties are satisfied &    \begin{minipage}[t]{2.8cm}
\texttt{len(l) = 5}
\end{minipage} \\ 
      Other                     & All other types of postconiditions &    \begin{minipage}[t]{2.8cm}
\texttt{isPrime(x)}
\end{minipage} \\ \bottomrule
  \end{tabular}
\end{table}

\section{Methodology}\label{sec:design}

\subsection{Maturity Model Design}

We introduce a maturity model based on postcondition generation to evaluate the maturity levels of code LLMs~(Table \ref{tab:maturity_model}).
This model provides deeper insights into the model capabilities of natural language understanding, logical reasoning, domain knowledge comprehension, and code generation. \looseness=-1


\begin{table}[tb]
    \centering
    \scriptsize
    \caption{Capabilities and Postcondition Types Mapping}
    \label{tab:maturity_model}
    \begin{tabular}{cllc}
        \toprule
        \textbf{Level} & \textbf{Capabilities} & \textbf{Postcondition Types} & \textbf{Proficiency Level} \\ \midrule
        Level 0 & \begin{tabular}[c]{@{}l@{}}
                   Natural Language Understanding \\ 
                   Logical Reasoning \\ 
                   Domain Knowledge Comprehension \\ 
                   Code Generation 
                  \end{tabular} & N/A & Low \\ \midrule
        Level 1 & \begin{tabular}[c]{@{}l@{}}
                   Basic Natural Language Understanding \\ 
                   Basic Logical Reasoning \\ 
                   Basic Domain Knowledge Comprehension \\ 
                   Basic Code Generation 
                  \end{tabular} & \begin{tabular}[c]{@{}l@{}}
                                    Type Checks \\ 
                                    NULL Checks 
                                   \end{tabular} & Low \\ \midrule
        Level 2 & \begin{tabular}[c]{@{}l@{}}
                   Intermediate Natural Language Understanding \\ 
                   Intermediate Logical Reasoning \\ 
                   Intermediate Domain Knowledge Comprehension \\ 
                   Intermediate Code Generation 
                  \end{tabular} & \begin{tabular}[c]{@{}l@{}}
                                    Boundary Checks \\ 
                                    Arithmetic Bounds \\ 
                                    Boolean Conditions \\ 
                                    Container Property  
                                   \end{tabular} & Medium \\ \midrule
        Level 3 & \begin{tabular}[c]{@{}l@{}}
                   Advanced Natural Language Understanding \\ 
                   Advanced Logical Reasoning \\ 
                   Advanced Domain Knowledge Comprehension \\ 
                   Advanced Code Generation 
                  \end{tabular} & \begin{tabular}[c]{@{}l@{}}
                                    Container Element \\ 
                                    Equality Checks 
                                   \end{tabular} & High \\ \midrule
        Level 4 & \begin{tabular}[c]{@{}l@{}}
                   Complete Natural Language Understanding \\ 
                   Complete Logical Reasoning \\ 
                   Complete Domain Knowledge Comprehension \\ 
                   Complete Code Generation 
                  \end{tabular} & All Types & High \\ \bottomrule
    \end{tabular}
\end{table}

\textbf{Level 0: Preliminary Understanding.}
The LLM lacks the fundamental understanding necessary for the postcondition generation task. Its natural language understanding is severely limited, leading to low accuracy in interpreting program descriptions or requirements. The LLM cannot infer logic, conditions, and outcomes of programs from descriptions, which are essential for generating postconditions. It has minimal or no knowledge of programming conventions or postcondition concepts, and its domain knowledge is inaccurate and incomplete. Additionally, the LLM is unable to generate coherent and expected code related to the descriptions.

\textbf{Level 1: Basic Postcondition Generation.}
The LLM can generate simple and straightforward postconditions that capture the most fundamental program states and properties. It demonstrates a basic understanding of natural language, sufficient to comprehend major program functionality described in descriptions and requirements. The LLM can identify the most basic and common properties or states of the program from descriptions and handle straightforward logical conditions. It has a rudimentary comprehension of programming conventions and postcondition concepts, allowing it to generate simple postconditions. The LLM also has limited domain knowledge related to the program and can generate simple and coherent code that includes basic conditional logic.

For our postcondition taxonomy, \textit{type} and \textit{NULL} postcondition generation tasks are suitable to evaluate whether LLMs reach Level 1. These tasks involve basic checks that require the LLM to identify variable types and null values, which are fundamental skills necessary at this level.

\textbf{Level 2: Comprehensive Postcondition Generation.}
The LLM can generate more comprehensive postconditions that reflect partial program logic and execution scenarios. It shows an enhanced understanding of natural language, with intermediate accuracy in interpreting the program's intention and behavior from descriptions. The LLM can infer partial logic, conditions, and outcomes of the program from descriptions. It has an enhanced comprehension of programming conventions, postcondition concepts, and specific domain knowledge related to the program. The LLM can generate more complex code related to the program, involving control flow, nested conditions, loops, and other advanced constructs.

\textit{Boundary checks}, \textit{arithmetic bounds}, \textit{Boolean conditions}, and \textit{container property} postconditions  are great for Level 2 evaluation. These tasks require the LLM to handle more complex conditions and ranges, reflecting an intermediate understanding of program logic and properties.

\textbf{Level 3: Advanced Postcondition Generation.}
The LLM can handle a wide range of postconditions that cover the program logic and execution scenarios, including edge cases and exceptions. It demonstrates an advanced understanding of natural language, with high accuracy in extracting the program's intention and behaviors from even more complex descriptions. The LLM can infer more complete and concrete logic, conditions, and outcomes of the program from descriptions. It has an advanced comprehension of programming conventions, postcondition concepts, and specific domain knowledge related to the program. The LLM is proficient in generating coherent code, dealing with complex data structures or logical implementations.

Postcondition generation tasks like \textit{container element} and \textit{equality} postconditions assess whether LLMs reach Level 3. 
These tasks involve verifying the correctness of elements within containers and ensuring equality, requiring a deeper understanding of data structures and precise logical reasoning.

\textbf{Level 4: Automated Postcondition Support.}
The LLM is able to support automated postcondition generation with absolute precision. It demonstrates complete understanding of natural language and is capable of extracting the correct program's intention and behaviors from descriptions. The LLM can infer complete logic, conditions, and outcomes of programs from descriptions. It has deep and broad comprehension of programming conventions, postcondition concepts, and specific domain knowledge related to the program. The LLM can generate accurate code across all postcondition types, achieving high proficiency in every aspect of postcondition generation.

Performance across all postcondition generation tasks determines whether LLMs are weak at Level 0 or perfect at Level 4. Achieving Level 4 indicates the LLM's ability to handle the full range of postcondition types with complete accuracy and precision.

\subsection{Prompt Generation}

\begin{figure}
    \centering
    \footnotesize
    \includegraphics[width=\linewidth]{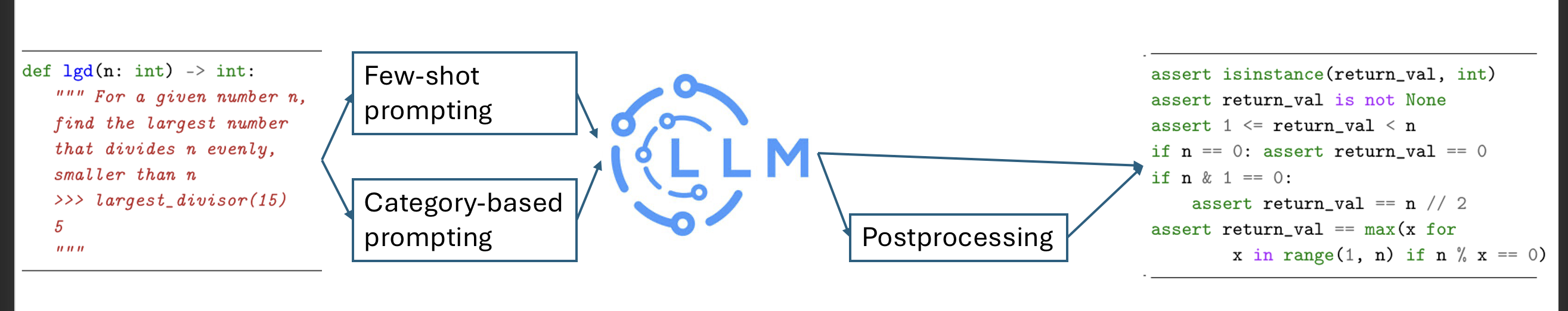}
    \caption{Overview of Prompt Generation }\label{fig:approach}
    \vspace{-10pt}
\end{figure}

The overview of our prompt generation approach is presented in \autoref{fig:approach}, which primarily consists of two generation logics: few-shot generation and category-based generation. 
In the few-shot generation, we generate postconditions for the targeted program using a basic prompt that includes a few examples.
This approach leverages the model’s ability to learn from a small number of examples to generate relevant postconditions for the given program. 
In the category-based generation, we decompose the postcondition generation task into multiple simpler subtasks based on specific categories. 
For each postcondition category, we use a more specific prompt that explicitly instructs the LLMs to generate postconditions of that particular category. 
This targeted prompting ensures that the model focuses on generating accurate and relevant postconditions for each category. 
In postprocessing phase, we exact and identify expected postconditions from LLM-generated results which ensures the correctness and relevance of the postconditions generated for each prompt. 
Finally, we gather all the postconditions generated from the few-shot generation and each subtask in the category-based generation.

\subsection{Few-short Generation}

Notice that most LLMs are not trained to generate postconditions. 
As such, we use the few-shot generation as the baseline method. 
The prompt consists of five components: an instruction, guidelines, detailed response format, few-shot examples, and a natural language description of the targeted program. 
We have manually experimented with many different prompts and chose the best one based on experts’ evaluation. 
The prompt is presented in \autoref{fig:prompt}.

The instruction describes the task for the LLMs, which is to generate postconditions in the form of assertions. 
The guidelines include basic rules for generating postconditions, which aims to improve readability and accuracy of the postconditions. 
The response format ensures that these postconditions can be easily extracted. 

Few-shot examples are included to enhance the LLMs’ performance and teach them the desired response format. 
These examples, written manually by researchers, include a natural language description of a program and the corresponding postconditions. 
The examples are designed to be simple, accurately describe the program’s intent and behaviors, be correct and consistent, and cover as many postcondition categories as possible to teach the models about different postcondition types.
Finally, the last part of the prompt is the natural language description of the targeted program. 
Details of the prompts are explained in \autoref{sec:template}.

\subsection{Category-based Generation}


Basic prompts may suffice for straightforward scenarios, but category-based prompts are crucial for capturing the nuances of different postcondition categories.
In the category-based generation, we break down the complex task of generating postconditions into multiple simpler subtasks, each focusing on generating a specific category of postconditions. 
Simpler but more specific tasks help LLMs understand our requirements better and produce higher-quality outputs. 
Additionally, multiple category-specific subtasks facilitate more effective evaluation of the model's capabilities across various postcondition types, ensuring a comprehensive assessment of its proficiency and robustness.

More specifically, we generate \textit{type}, \textit{NULL}, \textit{boundary}, and \textit{equality} postconditions for each program. 
Additionally, postconditions for \textit{arithmetic bounds}, \textit{boolean condition}, \textit{string format} and \textit{container elements or properties} check are generated selectively, based on the specific data type requirements of the variables in the program.

We leverage specialized prompt for each category, which also consists of five same components as basic prompt. However, these prompts provide clearer, more detailed instructions for specific category, improving the model's ability to generate accurate and relevant postconditions. The guidelines and few-shots examples parts varies according to different postcondition categories. Some rules in guidelines are more strict as the expected postconditions are more specific. For few-shot examples, we only keep the postconditions that belong to the expected category.

\section{Results}\label{sec:Eval}

\subsection{Experimental setups}



\textbf{Datasets:} We employed the EvalPlus dataset~\cite{Liu2023IsYC} 
which comprises 164 coding problems, each featuring essential code context, a function signature, descriptive comments, a canonical solution, and a comprehensive set of legal test inputs. 
These test sets are meticulously designed to validate the accuracy of LLM-generated code, which are good for assessing LLM-generated postconditions. 
Following the standard approach, we generated code mutants embedded with potential errors to further evaluate the reliability of these postconditions. 
Specifically, we utilized the Gemma-7b-it~\cite{Mesnard2024GemmaOM} and Mistral-7B-Instruct-v0.2~\cite{Jiang2023Mistral7} models at a temperature setting of 0.9 in the mutant generation. We directly prompted the models to introduce bugs into the code. 
Each problem underwent at least 50 iterations of bug insertion, retaining only those mutants that failed the test sets. 
Subsequently, we removed duplicates among the failing mutants that exhibited identical failures across the same test cases. 
In total, we compiled 1,293 buggy code mutants, averaging approximately 8 bugs per problem.




\textbf{Large Language Models:} We selected three state-of-the-art, open-source LLMs for generating postconditions based on natural language descriptions.

\noindent \textit{Gemma:} Both Gemma-7b-it and Gemma-1.1-7b-it are instruction-tuned versions of the Gemma-7b pre-trained model. 
The Gemma-1.1-7b-it model has undergone additional fine-tuning using Reinforcement Learning with Human Feedback (RLHF), which has enabled it to achieve state-of-the-art performance in coding tasks compared to other models of similar size~\cite{Mesnard2024GemmaOM}. 
Access to both Gemma models is provided through HuggingFace's APIs\cite{HuggingFace}.

\noindent \textit{Mistral:} The Mistral-7B-Instruct-v0.2 model is an instruction-tuned variant of the pre-trained Mistral 7B model. 
It has demonstrated exceptional performance across various tasks, particularly in reasoning, mathematics, and coding~\cite{Jiang2023Mistral7}. 
Access to this model is also facilitated via HuggingFace's APIs\cite{HuggingFace}.





\textbf{Metrics:} We employ four key metrics to evaluate the efficacy of LLMs in generating postconditions and to assess the quality of the generated postconditions.


\noindent \textit{Correct Postcondition Count (CPC)}: This metric quantifies the total number of problems for which the LLM has generated entirely correct postconditions.

\noindent \textit{Coverage@k (C@k)}: An extension of the traditional accept@k metric, this evaluates the probability of obtaining at least one correct postcondition in a sample of k responses. 
    It is calculated by examining subsets of responses from the model, from size 1 to m, to determine the expected value of obtaining at least one valid postcondition. 
    This metric helps assess the robustness of LLM performance across multiple trials and reduce the impact of the randomness of LLM-generated results.
    
\noindent \textit{Bug Detection Rate (BDR)}: This measures the percentage of buggy code mutants that are accurately identified by the postconditions. 
    It reflects the model's ability to pinpoint errors within code.

\noindent \textit{Bug Coverage Rate (BCR)}: This metric calculates the proportion of test problems where the LLM-generated postconditions collectively identify all present bugs, thus providing a measure of the comprehensiveness of the model’s error detection.

\textbf{Postcondition Generation:} For each problem in the EvalPlus dataset, we generated responses five times per prompt for each LLM model. 
The sampling temperature was set to 0.7 to optimize the balance between the diversity and precision of the outputs from the LLMs. 
For few-shot generation, we utilized zero-shot, one-shot, and three-shot scenarios to facilitate a comparative analysis across few-shot settings.
For category-based generation, we gather all the LLM-generated postconditions from each category as an overall result for a comparison to the few-shot generation.
Lastly, we also combine the results of 3-shots and category-based generation for further evaluation.
\subsection{Correctness of LLM-generated Postconditions}

\begin{table}[tb]
    \centering\scriptsize
    \caption{Correctness and Completeness of Generated Postconditions}\label{tab:correctnesscompleteness}
    \resizebox{\textwidth}{!}{%
    \begin{tabular}{l l r r r r r r}
        \toprule
        \textbf{Model} & \textbf{Approach} & \multicolumn{1}{c}{\textbf{C@1 (\%)}} & \multicolumn{1}{c}{\textbf{C@3 (\%)}} & \multicolumn{1}{c}{\textbf{C@5 (\%)}} & \multicolumn{1}{c}{\textbf{CPC}} & \textbf{BCR(\%)} & \textbf{BDR(\%)} \\
        \midrule
        \multirow{5}{*}{Gemma-7b-it} & 0-shot & 47.44 & 65.73 & 71.34 & 117 & 10.37 & 27.84 \\
                                     & 1-shot & 98.54 & 99.94 & 100.00 & \textbf{164} & 18.90 & 50.19 \\
                                     & 3-shots & 99.76 & 100.00 & 100.00 & \textbf{164} & 15.24 & 46.71 \\
                                     & Category-based & {--} & {--} & {--} & \textbf{164} & 18.90 & 60.23 \\
                                     & \makecell[l]{3-shots+\\Category-based} & {--} & {--} & {--} & \textbf{164} & \textbf{21.34} &\textbf{63.34} \\
        \midrule
        \multirow{5}{*}{Gemma-v1.1-7b-it} & 0-shot & 69.15 & 83.35 & 87.20 & 143 & 11.59 & 38.59 \\
                                          & 1-shot & 92.44 & 96.40 & 97.56 & 160 & 23.17 & 58.86 \\
                                          & 3-shots & 99.88 & 100.00 & 100.00 & \textbf{164} & 21.34 & 55.38 \\
                                          & Category-based & {--} & {--} & {--} & \textbf{164} & 31.72 & 72.31 \\
                                          & \makecell[l]{3-shots+\\Category-based} & {--} & {--} & {--} & \textbf{164} & \textbf{34.15} & \textbf{75.72} \\
        \midrule
        \multirow{5}{*}{Mistral-7B-Instruct} & 0-shot & 69.27 & 89.27 & 92.68 & 152 & 15.24 & 44.62 \\
                                             & 1-shot & 90.61 & 97.26 & 98.78 & 162 & 23.78 & 53.52 \\
                                             & 3-shots & 98.41 & 100.00 & 100.00 & \textbf{164} & 25.61 & 54.52 \\
                                             & Category-based & {--} & {--} & {--} & \textbf{164} & 29.88 & 75.56 \\
                                             & \makecell[l]{3-shots+\\Category-based} & {--} & {--} & {--} & \textbf{164} & \textbf{31.10} &\textbf{77.42} \\
        \bottomrule
    \end{tabular}
}
\end{table}

From \autoref{tab:correctnesscompleteness}, we observe that in all three models, the 3-shots approach and the category-based approach generated at least one correct postcondition for each problem in EvalPlus, significantly outperforming the 0-shot approach with only 117 problems solved when using Gemma-7b-it model and slightly better than the 1-shot approach with over 160 problems solved. 
The C@k metric, especially C@1, increases as the number of few-shot examples increases, indicating that LLMs are more likely to generate valid responses in few-shot settings. 
This result highlights the effectiveness of our few-shot examples. 
The combination the 3-shots and the category-based approach achieves the same performance as the two individual approaches, generating correct postconditions for all the problems. \looseness=-1

Comparing LLM-generated postconditions among the 0-shot, few-shot approaches (both 1-shot and 3-shots), and the category-based approach, we find that the few-shot approaches generated a larger amount and more diverse types of postconditions than the 0-shot approach. 
For each model, we gathered over 3,000 postconditions in total from all responses in the few-shot approaches, whereas in the 0-shot approach, the number of generated postconditions was less than 2,000. 
This higher quantity likely contributed to more correct postconditions. 
Additionally, postconditions from the few-shot approaches covered more postcondition types, especially \textit{type}, \textit{NULL} checks, and \textit{boundary case} checks. 
This diversity in postcondition types enhances the likelihood of generating correct postconditions in LLM generations based on the results in \autoref{subsec:catogory}.

Under the 0-shot settings in basic generation, the gemma-v1.1-7b-it (\(C@1 = 69.15\%\)) and Mistral-7B-Instruct models (\(C@1 = 69.27\)) are more likely to generate valid responses containing correct postconditions, generally outperforming the gemma-7b-it model with C@1 being 47.44\%. 
But the C@5 of gemma-7b-it model is relatively high as 77.34\%, meaning that 77.34\% of problems in \sys have a valid response. 
As for few-shot settings(both 1-shot and 3-shots), all three models perform very well, with C@1 always higher than 90\%.

\subsection{LLMs Capabilities on Generating Different Types of Postconditions}\label{subsec:catogory}

\begin{table}[]
	\centering\footnotesize
    \caption{Correctness Performance on Each Postcondition Category of Gemma-v1.1-7b-it. \#Prob: the number of problems for which we generated postconditions, varies according to the category}
	\label{tab:category-correct}
	\begin{tabular}{lrrrrr}
		\toprule
		Category                 & C@1(\%) & C@3(\%) & C@5(\%) & \#Prob & CPC \\ 
        \midrule
		Type               & \textbf{97.20} & 98.84 & 99.39 & 164 & 163 \\
		NULL               & 95.37 & 97.38 & 97.56 & 164 & 160 \\
		Boundary      & 94.88 & \textbf{99.51} & \textbf{100.00\%} & 164 & 164 \\
		Equality           & 22.20 & 30.00 & 34.15 & 164 & 56 \\
		Arithmetic Bounds  & 79.29 & 85.54 & 87.50 & 56 & 49 \\
		Boolean Conditions          & 41.43 & 64.64 & 78.57\% & 28 & 22 \\
		String Format             & 54.00 & 77.33 & 83.33\% & 30 & 25 \\
		Container Elements & 50.00 & 67.60 & 72.00 & 50 & 36 \\
		Container Property & 70.80 & 88.20 & 94.00 & 50 & 47 \\ 
        \midrule
		Overall                    &  &  & & 164 & 164 \\ 
        \bottomrule
	\end{tabular}
\end{table}

We present the correctness results of the gemma-v1.1-7b-it model on individual types of postconditions in \autoref{tab:category-correct} for further analysis. 
The data of the other two models is presented in \autoref{sec:result}.
According to \autoref{tab:category-correct}, the gemma-v1.1-7b-it model exhibit significant variations in their correctness when generating different categories of postconditions. 
Specifically, for the three types of postconditions: \textit{type}, \textit{NULL}, and \textit{boundary} checks, the gemma-v1.1-7b-it model performs exceptionally well, with all three C@k metrics above 90\%. 
Particularly for \textit{boundary} checks, the model generated correct postconditions for each problem in EvalPlus. 
In addition, for \textit{arithmetic bounds} and \textit{container properties}, the model also achieved a high C@1 over 70\% and C@5 near 90\%. 
This indicates the model's high efficiency and accuracy in generating these types of postconditions.
However, for \textit{boolean conditions}, \textit{string formats}, and \textit{container elements} checks, the model performs poorly in C@1 around 50\%, but achieves relatively high C@5, all above 70\%. 
This suggests that the model may need multiple attempts to generate correct postconditions for these categories.
Lastly, the model's performance is worst for the \textit{equality} type of postconditions, with C@k ranging from 22.20\% to 34.15\%, which suggests that the LLM are unlikely to generate correct postconditions equality check.

These results indicate the model are sufficient for the level 1 of our maturity model, with a high accuracy in generating \textit{type} and \textit{NULL} check postcndition. 
Also we believe that the capabilities of the model have reach the second maturity level, proficiently generating postconditions for \textit{boundary}, \textit{arithmetic bounds}, \textit{boolean conditions}, and \textit{container property}. 
However, the model's performance on \textit{container elements} and \textit{equality} postcondition are relatively poor, which are related to level 4 of the maturity model.
Therefore, the model is considered in the second maturity level.

\subsection{Completeness of LLM-generated Postconditions}

From \autoref{tab:correctnesscompleteness}, we find that the approach combining 3-shots and category-based approach has the best performance among all the approaches, with achieving highest BCR as 34.15\% in Gemma-v1.1-7b-it model and highest BDR as 77.42\% in Mistral-7B-Instruct model.
This indicates that postconditions generated by LLMs can effectively detect bugs. 
Additionally, category-based approach with a BCR ranging from 18.90\% to 31.72\% and a BDR ranging from 60.23\% to 77.42\%, outperforms all few-shot approaches ($ 10.37\% \leq BCR \leq 25.61\% $, $ 27.84\% \leq BDR \leq 58.86\% $) in all three models . 
This is also due to the larger amount and more diverse types of correct postconditions generated by category-based approach, as each postcondition detects a certain amount of bugs.
Especially, \textit{boundary} and \textit{equality} check are more common in the results of category-based approach, which generally detects more bugs than the other categories.
Lastly, both one-shot($ 18.90\% \leq BCR \leq 23.78\% $, $ 50.19\% \leq BDR \leq 58.86\% $) and three-shots($ 15.24\% \leq BCR \leq 31.72\% $, $ 46.71\% \leq BDR \leq 55.38\% $) approaches outperform zero-shot($ 10.37\% \leq BCR \leq 15.24\% $, $ 27.84\% \leq BDR \leq 44.62\% $) approach in both metrics significantly, which further prove the effectiveness of our few-shot examples.

As for different models, Gemma-v1.1-7b-it and Mistral-7B-Instruct perform relatively well in our approach, followed by Gemma-7b-it model. 
Both Gemma-v1.1-7b-it and Mistral-7B-Instruct detected all bugs for over 30\% problems in EvalPlus while Gemma-7b-it model only solved 18.90\% problems. 
Similarly, the first two models detected 75\% of all bugs, outperforming Gemma-7b-it model with 63.34\% detected bugs. 
These results also indicate that the Gemma-v1.1-7b-it and Mistral-7B-Instruct models generated postconditions that can effectively formalize programs' logic, which reach the second level of our maturity model. 
As for Gemma-7b-it model, its generated postconditions can only cover part of program logic with a relatively low bug coverage rate(18.90\%) but medium high bug detection rate(63.34\%). So this model is considered to be the level 1 of the maturity model.

\subsection{Impacts of Individual Postconditioin Types}\label{sec:individual}

We have also evaluated the model on individual types of postconditions, and the detailed numbers are presented in \autoref{sec:completeness} and \autoref{sec:result}.
In summary, the results show that all models become less effective when the type of postcondition requires more advanced capabilities or a higher level of maturity in specific capabilities. 
These findings underscore the importance of developing a maturity model to accurately assess the quality of code LLMs.
\section{Conclusion}\label{sec:conclusion}
In this paper, we have introduced a novel maturity model for evaluating the capabilities of code LLMs through the lens of postcondition generation. 
By incorporating diverse postcondition types and augmenting the EvalPlus dataset, we enable a detailed evaluation of LLM performance across multiple dimensions of code understanding and generation. 
Our extensive evaluation of open-sourced models underscores the importance of advanced capabilities in generating correct and diverse postconditions, highlighting areas for improvement in current LLMs. 
Our work not only contributes to the development of more proficient and versatile code LLMs but also sets a new standard for the evaluation of AI models in software development, ultimately enhancing productivity and ensuring the responsible use of AI technology.

\section*{Limitations, Availability, Ethics, and Broader Impacts}\label{sec:discussion}

Limitations: \protect\circled{1} our maturity model incorporates a diverse set of postcondition types but may not cover all possible scenarios and edge cases encountered in real-world programming tasks, limiting generalizability; 
\protect\circled{2} the approach relies heavily on few-shot examples, and the quality and diversity of these examples significantly influence model performance, potentially leading to suboptimal results if examples are inadequate; 
and \protect\circled{3} evaluations are limited to a selection of open-sourced LLMs, and proprietary models with advanced training techniques and larger datasets might perform differently. 
To mitigate potential harms and boost productivity, we have open-sourced our benchmark. 
By making our benchmark available to the public, we aim to foster transparency, encourage collaboration, and facilitate further research and development. 

Our work has several broader impacts on both the research community and practical applications. 
\protect\circled{1} By introducing a comprehensive maturity model for evaluating code LLMs, we provide a more nuanced and detailed framework for assessing the capabilities of these models, improving their design and training. 
\protect\circled{2} The augmentation of the EvalPlus dataset to include postcondition testing sets a new standard for benchmark datasets, encouraging the development of more robust and versatile LLMs. 
\protect\circled{3} Open-sourcing our benchmark promotes transparency and collaboration within the AI and software development communities, enabling building upon our work and address any identified limitations or biases. 
\protect\circled{4} By facilitating the generation of more accurate and reliable code, our approach can significantly enhance productivity and reduce errors in software development, leading to more efficient and effective coding practices. 

\bibliographystyle{unsrtnat}
\bibliography{src/ref}

\newpage
\appendix

\section{Prompt Templates}\label{sec:template}

\begin{figure}[htbp]
    \centering
    \includegraphics[width=1\linewidth]{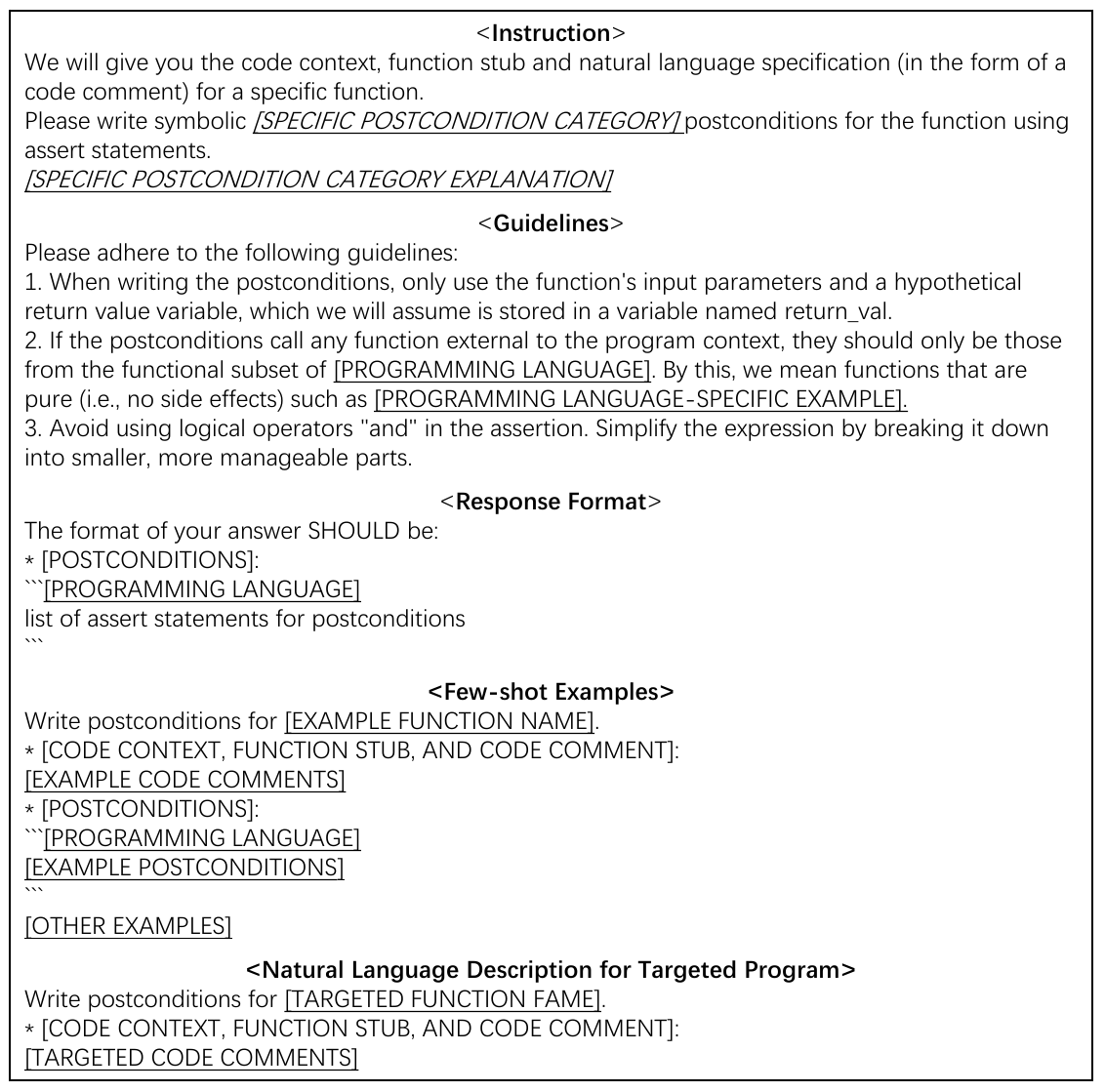}
    \caption{Prompt template for our approach. \underline{Underline} text would be replaced by concrete contents specific to the programming language, targeted program and postcondition category that we are handling. \textit{Italicized} text would only be included in category-specific prompts.}
    \label{fig:prompt} 
\end{figure}

\autoref{fig:prompt} shows the prompt we use to perform the few-shot prompting.

The instruction aims to describe our task to the LLMs, which is to generate postconditions in the form of assertions. 
The response format specifies the required format of LLMs’ responses, which helps us to extract postconditions from them.

The guidelines demonstrate some basic rules for generating postconditions. 
The first two rules were included in the original prompt. 
The third rule requires avoiding complex postconditions and breaking them down into atomic postconditions. 
An atomic postcondition should check only one specific aspect of the program. 
However, postconditions generated by LLMs usually consist of multiple atomic postconditions conjoined using \&\& (logical AND). 
If any atomic postcondition within a postcondition fails the test, the whole postcondition would be considered incorrect, even though the other atomic postconditions are correct. 
Thus, adopting the third rule not only helps generate simpler and more readable postconditions but also extracts possible correct atomic ones from complex postconditions.

Furthermore, we add some few-shot examples to the prompt. 
The reasons are twofold. 
First, few-shot examples can help improve the LLMs’ performance in various tasks \cite{Brown2020LanguageMA}. 
Second, LLMs can learn how to respond in the desired format from these examples. 
All few-shot examples are written manually by researchers and consist of two parts: a natural language description of a program and the corresponding postconditions.

When constructing the few-shot examples, we adhere to the following guidelines: \protect\circled{1} the program’s functionality should be relatively simple; 
\protect\circled{2} the corresponding description needs to convey the program’s intent and behaviors accurately and clearly;
\protect\circled{3} the postconditions should be correct, consistent with the program’s description, and complete enough to cover all the important aspects of the program; and \protect\circled{4} the postconditions are expected to cover as many postcondition categories as possible. 
We hope that models can learn the characteristics of different categories of postconditions and cover diverse categories, as each category checks different aspects of programs.

The last part of the prompt is the natural language description of the targeted program.

\section{Refining Answers}\label{sec:refine}

Even though we specify the expected postcondition category in our prompt, it is still possible for LLMs to generate postconditions that do not belong to that category. To adjust the results for later analysis, we design a strategy to filter out such postconditions. First, we list some common key words for each postcondition category, which is shown in \autoref{tab:keywords}. we select out the psotconditions assertions that matches with the keywords of expected category. Then, we conduct manual judgment on the results to make them more in line with our expectations. 

\begin{table}
    \centering
    \footnotesize
    \caption{Keywords for Each Postcondition Category}\label{tab:keywords}
    \begin{tabular}{rl}
		\toprule
		Category            & Keywords \\ \midrule
		Type                & \texttt{isinstance, type} \\ 
		NULL                & \texttt{None}  \\ 
		Boundary            & \texttt{if, ==, is} \\ 
		Equality            & \texttt{is, ==} \\ 
		Arithmetic Bounds   & \texttt{<, >, <=, >=, in} \\ 
		Boolean Condition   & \texttt{if, is, ==, True, False}\\ 
		String Format       & \texttt{for...in..., in, len, startswith, ...} \\ 
		Container Element   & \texttt{for...in..., array[...], ...} \\ 
		Container Property  & \texttt{len, sum, count, ...} \\ 
        \bottomrule
    \end{tabular}
\end{table}

\section{Contributions of Individual Category in Bug Detection}\label{sec:completeness}

\begin{table}
	\centering
    \caption{Completeness on Postcondition Categories of Gemma-v1.1-7b-it. \#Prob: the number of problems for which we generated postconditions. \#Bugs: the total number of bugs.}\label{tab:category-complete}
	\begin{tabular}{lrrrrrr}
 	\toprule
		Category           & \#Prob & BCC & BCR(\%)  & \#Bugs & BDC & BDR(\%) \\
        \midrule
		Type               & 164      & 5            & 3.05         & 1293 & 273         & 21.11      \\
		NULL               & 164      & 2            & 1.22         & 1293 & 155         & 11.99      \\
		Boundary      & 164      & 25           & 15.24        & 1293 & \textbf{743}& \textbf{57.46}      \\
		Equality           & 164      & \textbf{34}  & \textbf{20.73} & 1293 & 263       & 20.34      \\
		Arithmetic Bounds  & 56       & 5            & 8.93         & 376  & 144         & 38.30      \\
		Boolean Conditions & 28       & 2            & 7.14         & 234  & 103         & 44.02      \\
		String Format      & 30       & 1            & 3.33         & 279  & 116         & 41.58      \\
		Container Elements & 50       & 4            & 8.00         & 404  & 140         & 34.65      \\
		Container Property & 50       & 0            & 0.00         & 404  & 174         & 43.07      \\ 
        \midrule
		Overall            & 164      & 52           & 31.71\%        & 1293 & 935         & 72.31   \\ 
        \bottomrule   
	\end{tabular}
\end{table}


From \autoref{tab:category-complete}, we find that bug detection capability of different categories of LLM-generated postconditions also varies significantly. 
\textit{Bounds} Check achieve relatively high in both BCR and BDR metrics. 
This suggest that it captures important behaviors of program. 
\textit{Equality} check have a highest BCR as 20.73\%, but a rather low BDR. This might be caused by the poor performance in correctness.
In contrast, other categories like \textit{Arithemetic Bounds}, \textit{Boolean Conditions}, \textit{String Format}, \textit{Container Elements/property}, have a low BCR but a relatively high BDR. 
Such postcoditions might not cover all the behaviors and states, but verify the important ones in some degree.
Lastly, \textit{type} and \textit{Null} check work poorly in both metrics, as they only focus on some small properties of programs.

Different categories of postconditions focus on various aspects of a program, thus leading to significant differences in the types and quantity of bugs they can detect. 
For some simpler postcondition categories, such as \textit{type} and \textit{NULL} checking, their focus is relatively narrow, targeting specific attributes and states within the program. 
In contrast, other postconditions, like \textit{equality} and \textit{boundary} check, are more capable of reflecting the complex logic and exceptional situations within the program, often uncovering more potential issues during defect detection.
This also confirms our rational of different levels of the maturity model.


\section{Category-based Generation Results of Gemma-7b-it and Mistral-7B-Instruct}\label{sec:result}
\begin{table}[]
	\centering
    \scriptsize
    \caption{Correctness Performance on Each Postcondition Category of Gemma-7b-it. \#Prob: the number of problems for which we generated postconditions, varies according to the category}
	\begin{tabular}{lrrrrr}
		\toprule
		Category                 & C@1 & C@3 & C@5 & \#Prob & CPC \\ 
        \midrule
		Type               & 94.27 & 98.23 & 98.78 & 164 & 162 \\
		NULL               & 97.68 & 98.17 & 98.17 & 164 & 161 \\
		Boundary           & 77.80 & 94.82 & 98.17 & 164 & 161 \\
		Equality           & 18.04 & 27.20 & 33.54 & 164 & 55 \\
		Arithmetic Bounds  & 79.29 & 84.64 & 85.71 & 56 & 48 \\
		Boolean Conditions & 56.43 & 78.93 & 85.71 & 28 & 24 \\
		String Format      & 60.00 & 80.67 & 83.33 & 30 & 25 \\
		Container Elements & 40.71 & 58.40 & 66.00 50 & 33 \\
		Container Property & 62.00 & 76.60 & 80.00 50 & 40 \\ 
        \midrule
		Overall                    &  &  & & 164 & 164 \\ 
        \bottomrule
	\end{tabular}
\end{table}

\begin{table}
	\centering
    \scriptsize
    \caption{Completeness on Postcondition Categories of Gemma-7b-it. \#Prob: the number of problems for which we generated postconditions. \#Bugs: the total number of bugs.}
	\begin{tabular}{lrrrrrr}
 	\toprule
		Category           & \#Prob & BCC & BCR(\%)  & \#Bugs & BDC & BDR(\%) \\
        \midrule
		Type               & 164      &  4  & 2.44 & 1293 & 236          & 18.22      \\
		NULL               & 164      &  2           & 1.22         & 1293 & 149          &   11.51    \\
		Boundary      & 164      &  12          & 7.32        & 1293 & 474 & 36.60     \\
		Equality           & 164      & 16  & 9.76 & 1293 & 186        &  14.36     \\
		Arithmetic Bounds  & 56       & 4            &  7.14        & 376  & 116          & 30.85      \\
		Boolean Conditions & 28       & 4            & 14.29         & 234  & 113          & 47.88      \\
		String Format      & 30       & 2            &  6.67        & 279  & 96          & 34.41      \\
		Container Elements & 50       & 2            &  4.00        & 404  & 107          & 26.49      \\
		Container Property & 50       & 1            &  2.00        & 404  & 131          & 32.43      \\ 
        \midrule
		Overall            & 164      & 31          &  18.90       & 1293 & 780          & 60.23  \\ 
        \bottomrule   
	\end{tabular}
\end{table}

\begin{table}[]
	\centering
    \scriptsize
    \caption{Correctness Performance on Each Postcondition Category of Mistral-7B-Instruct. \#Prob: the number of problems for which we generated postconditions, varies according to the category}
	\begin{tabular}{lrrrrr}
		\toprule
		Category                 & C@1(\%) & C@3 & C@5 & \#Prob & CPC \\ 
        \midrule
		Type               & 97.20 & 98.54 & 98.78 & 164 & 162 \\
		NULL               & 89.51 & 97.68 & 98.17 & 164 & 161 \\
		Boundary      & 84.15 & 94.39 & 96.34 & 164 & 158 \\
		Equality           & 34.51 & 50.24 & 56.71 & 164 & 93 \\
		Arithmetic Bounds  & 81.79 & 92.14 & 94.64 & 56 & 53 \\
		Boolean Conditions          & 73.57 & 93.21 & 96.43 & 28 & 27 \\
		String Format             & 62.67 & 83.33 & 90.00 & 30 & 27 \\
		Container Elements & 63.60 & 85.60 & 94.00 & 50 & 47 \\
		Container Property & 90.40 & 95.60 & 96.00 & 50 & 48 \\ 
        \midrule
		Overall                    &  &  & & 164 & 164 \\ 
        \bottomrule
	\end{tabular}
\end{table}

\begin{table}
	\centering
    \scriptsize
    \caption{Completeness on Postcondition Categories of Mistral-7B-Instruct. \#Prob: the number of problems for which we generated postconditions. \#Bugs: the total number of bugs.}
	\begin{tabular}{lrrrrrr}
 	\toprule
		Category           & \#Prob & BCC & BCR(\%)  & \#Bugs & BDC & BDR(\%) \\
        \midrule
		Type               & 164      &  3           & 1.83         & 1293 & 221          & 17.09      \\
		NULL               & 164      &  5           & 3.05         & 1293 & 235          &   18.17    \\
		Boundary      & 164      &  31          & 18.90        & 1293 & 719 & 55.61     \\
		Equality           & 164      & 33  & 20.12 & 1293 & 500        &  38.67     \\
		Arithmetic Bounds  & 56       & 7            &  12.50        & 376  & 153          & 40.69      \\
		Boolean Conditions & 28       & 10            & 35.71         & 234  & 151          & 64.53      \\
		String Format      & 30       & 2            &  6.67        & 279  & 114          & 40.86      \\
		Container Elements & 50       & 3            &  6.00        & 404  & 180          & 44.55      \\
		Container Property & 50       & 1            &  2.00        & 404  & 222          & 54.95      \\ 
        \midrule
		Overall            & 164      & 49           &  29.88       & 1293 & 977          & 75.56   \\ 
        \bottomrule   
	\end{tabular}
\end{table}

\end{document}